\documentclass[a4paper,12pt]{article}
\usepackage{epsfig}
\usepackage{amssymb}
\usepackage{amsfonts}
\usepackage{graphics}
\usepackage{color,amsmath,bm}

\newskip\humongous \humongous=0pt plus 1000pt minus 1000pt

\newif\ifdtup

\catcode`\@=11

\@addtoreset{equation}{section}

\def\@normalsize{\@setsize\normalsize{15pt}\xiipt\@xiipt
\abovedisplayskip 14pt plus3pt minus3pt%
\belowdisplayskip \abovedisplayskip
\abovedisplayshortskip \z@ plus3pt%
\belowdisplayshortskip 7pt plus3.5pt minus0pt}

\def\small{\@setsize\small{13.6pt}\xipt\@xipt
\abovedisplayskip 13pt plus3pt minus3pt%
\belowdisplayskip \abovedisplayskip
\abovedisplayshortskip \z@ plus3pt%
\belowdisplayshortskip 7pt plus3.5pt minus0pt
\def\@listi{\parsep 4.5pt plus 2pt minus 1pt
     \itemsep \parsep
     \topsep 9pt plus 3pt minus 3pt}}

\relax

\catcode`@=12

\catcode`\@=11

\def\section{\@startsection{section}{1}{\z@}{3.5ex plus 1ex minus
   .2ex}{2.3ex plus .2ex}{\large\bf}}

\def\thesection{\arabic{section}}

\def\appendix{\setcounter{section}{0}
\def\thesection{\Alph{section}}}

\begin{document}

\newcommand{\beq}{\begin{equation}}
\newcommand{\eeq}{\end{equation}}
\newcommand{\bea}{\begin{eqnarray}}
\newcommand{\eea}{\end{eqnarray}}
\newcommand{\beas}{\begin{eqnarray*}}
\newcommand{\eeas}{\end{eqnarray*}}
\newcommand{\defi}{\stackrel{\rm def}{=}}
\newcommand{\non}{\nonumber}
\newcommand{\bquo}{\begin{quote}}
\newcommand{\enqu}{\end{quote}}
\newcommand{\mat}{\mathbf}
\def\d{\mathrm{d}}
\def\de{\partial}
\def\const{\hbox {\rm const.}}
\def\o{\over}
\def\im{\hbox{\rm Im}}
\def\re{\hbox{\rm Re}}
\def\bra{\langle}\def\ket{\rangle}
\def\Arg{\hbox {\rm Arg}}
\def\Re{\hbox {\rm Re}}
\def\Im{\hbox {\rm Im}}
\def\diag{\hbox{\rm diag}}
\def\longvert{{\rule[-2mm]{0.1mm}{7mm}}\,}
\def\Z{\mathbb Z}
\def\N{{\cal N}}
\def\tq{{\widetilde q}}
\def\W{{\cal W}}
\def\tQ{{\widetilde Q}}
\def\dag{{}^{\dagger}}
\def\p{{}^{\,\prime}}
\def\a{\alpha}
\def\Tr{ \hbox{\rm Tr\,}}
\def\tM{{\widetilde M}}
\def\tPhi{{\widetilde \Phi}}
\def\tphi{{\widetilde \phi}}
\def\tm{{\widetilde m}}
\def\tv{{\widetilde v}}
\def\tl{{\widetilde \lambda}}
\def\tg{{\widetilde g}}
\def\T{{\cal T}}
\def\t{T}
\def\J{{\cal J}}
\def\V{{\sf V}}
\def\lcm{\mathrm{lcm}}
\def\L{{\cal L}}

\begin{titlepage}

\renewcommand{\thefootnote}{\fnsymbol{footnote}}
\bigskip

\begin{center}
{\Large  {\bf Large $N$, $\mathbb{Z}_N$ Strings and Bag Models }
 }
\end{center}

\renewcommand{\thefootnote}{\fnsymbol{footnote}}
\bigskip
\begin{center}
{\large   Stefano Bolognesi }
 \vskip 0.20cm
\end{center}

\begin{center}
{\it      \footnotesize
Scuola Normale Superiore - Pisa, Piazza dei Cavalieri 7, Pisa, Italy \\
\vskip 0.10cm
and\\
\vskip 0.10cm
The Niels Bohr Institute, Blegdamsvej 17, DK/2100 Copenhagen, Denmark\footnote{Permanent address}} \\
\end {center}

\renewcommand{\thefootnote}{\arabic{footnote}}

\setcounter{footnote}{0}

\bigskip
\bigskip

\noindent
\begin{center} {\bf Abstract} \end{center}

We study $\mathbb{Z}_N$ strings in nonabelian gauge theories, when
they can be considered as domain walls compactified on a cylinder
and stabilized by the flux inside. To make the wall vortex
approximation reliable, we must take the 't Hooft large $N$ limit.
Our construction has many points in common with the
phenomenological bag models of hadrons.

\vfill

\begin{flushleft}
September, 2005
\end{flushleft}

\end{titlepage}

\bigskip

\hfill{}

\tableofcontents

\section{Introduction}

In a previous paper \cite{wallandfluxes}, we explored the idea
that a flux tube can be though, under particular circumstances, as
a domain wall compactified on a cylinder. To make it possible, the
theory must admit two degenerate vacua: one in the Coulomb phase,
that is kept inside the cylinder, and the other one in the Higgs
(or confining) phase, that is kept outside the cylinder.  For the
simplest example, the abelian Higgs-Coulomb model, we were able to
determine a condition under which the wall vortex approximation is
quantitative reliable. When we increase the number $n$ of quanta
of magnetic flux, the radius of the vortex grows like
$n^{\frac{2}{3}}$, while the thickness of the wall remains fixed.
When $n$ is enough large, the thickness of the wall becomes
negligible and the tension of the vortex is given by the simple
minimization of the energy density.

The present paper started with the following question: can we find
a nonabelian version of the wall vortex? In \cite{wallandfluxes}
we found examples where $\mathbb{Z}_N$ strings in a confining
$SU(N)$ gauge theory can be qualitative thought as wall vortices.
What we are locking for, is some realization in which the wall
vortex in quantitative reliable, so that we can apply the simple
argument of the energy minimization used for the abelian
Higgs-Coulomb model.

If we want to mimic what we obtained for the abelian model, we
must increase the number $n$ of flux quanta, until the radius of
the $n$-string becomes much larger than the thickness of the wall.
A problem immediately arise: in a $SU(N)$ gauge theory, $n$ is
limited to be smaller than $N$. This means that, to have a chance
of obtaining some result, we must explore the large $N$ limit of
the nonabelian gauge theory \cite{planarthooft}.

In the first part of the paper we will explore the case of
degenerate vacua, one in the Coulomb phase and the other one in
the confining phase,  and we will consider two examples of
supersymmetric gauge theories. In the second part of the paper we
face the more general situation in which the Coulomb vacuum is
metastable (or in the extreme case instable). Our hope is to apply
these  ideas to non supersymmetric gauge theories and we will try
to do that for large $N$ QCD.

The ideas we will expose are not new to particle physics. Almost
thirty years ago a lot of works have been made on the bag models
of hadrons. The first one was the so called MIT bag model
\cite{MITbag}, where the quarks where modelled as free fermions
bounded in region of finite volume and non zero energy density.
This model had a great success in explaining static properties of
hadrons \cite{DeGrand:1975cf}. A lot of different version of the
bag model have been proposed after that. For example, in the so
called SLAC bag model \cite{SLACbag}, the bag had finite tension
and the interior of the bag had zero energy density. Another
interesting approach was the Friedberg-Lee model of hadrons
\cite{Friedberg:1977xf}.  In this works, by means of an auxiliary
scalar field, they derived the bag as a domain wall interpolating
between a metastable vacuum and a true vacuum. In the metastable
vacuum the quarks have a small mass, while in the true vacuum they
have a great (or infinite) mass. In the Friedberg-Lee model,
hadrons where realized as nontopological solitons.  Despite their
success in explaining static properties of hadrons, bag model had
some difficulties in describing interactions and so they where
though only as phenomenological modelling of the real theory, QCD.
Attempts to derive the bag model from QCD first principles have
been made in \cite{fr} and \cite{gross}.

\begin{figure}[h!t]
\begin{center}
\leavevmode \epsfxsize 10 cm \epsffile{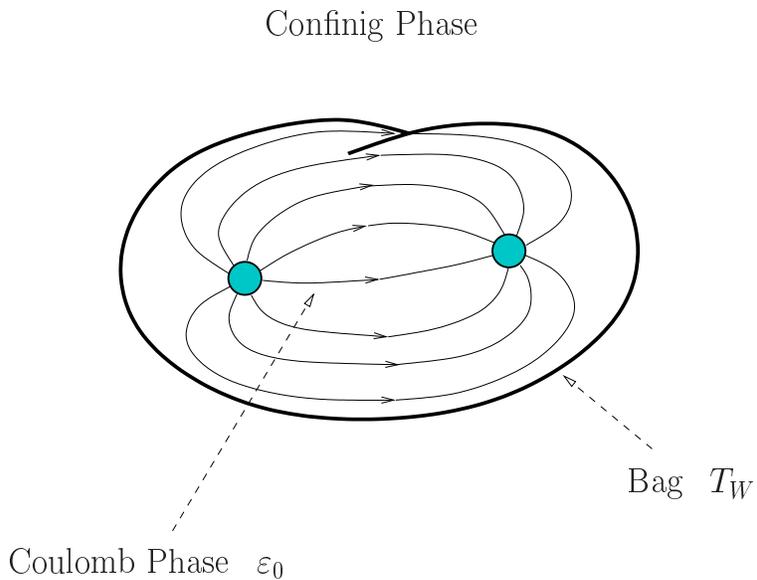}
\end{center}
\caption{\footnotesize  A meson in the bag model. Quarks and gauge
fields are in a Coulomb phase trapped inside the bag. The Coulomb
phase has energy density $\varepsilon_0$ and the bag has surface
tension $T_W$. The confining vacuum is outside the bag.}
\label{bag}
\end{figure}

In this paper we will argue that some confining gauge theory can
share, in the large $N$ limit, some properties in common with bag
models. To be more clear, we give now a definition of what we mean
by a bag model. The definition is not precise but sufficiently
large to include all the possible realizations. In a bag model we
have two phases: one in the interior of the bag that can contain
quarks and gluons, the other in the exterior of the bag where only
gauge singlets can live.  The interior Coulomb phase can have some
energy density $\varepsilon_0$ and the bag can have a tension
$T_W$. For example, in the MIT bag model $\varepsilon_0 \neq 0$
and $T_W =0$, while in the SLAC bag model $\varepsilon_0=0$ and
$T_W \neq 0$. In general we can define a bag model action: \beq L=
\int_V \d^3 x \, \L_{int} - \varepsilon_0 V -T_W \int_S \d^2 \xi
\, \sqrt{-\det{h}} \ ,\eeq where $\L_{int}$ describes the dynamics
in the interior Coulomb phase. Another important ingredient, to
complete the definition of the theory, consist in specifying the
boundary conditions of the fields at the bag surface. This
condition is that color must be trapped inside the bag. In Figure
\ref{bag} there is a meson in the bag model. When the meson is
rotating very fast, it becomes approximatively the wall vortex
with a quark and an antiquark at the ends. The connection between
the bag and the string was first recognized in
\cite{Johnson:1975sg}.

In the large $N$ limit of a $SU(N)$ gauge theory, Feymann graphs
organize themselves into a genus expansion in powers if
$\frac{1}{N}$ \cite{planarthooft}.  For this reason it is believed
that some dual string theory should describe nonabelian gauge
theories, and this string theory should become weakly coupled as
$N$ goes to infinity. The AdS/CFT correspondence is a concrete
realization of this duality \cite{Maldacenaconjecture} for the
$\N=4$ superconformal gauge theory. After that, some examples have
been found where the gauge theory is not superconformal but
confining \cite{PolchinskiStrassler,KlebanovStrassler}.  For the
above mentioned reasons it may be sound a bit strange that a
confining gauge theory becomes well approximated by a bag model in
the large $N$ limit.
An important point, is that our claims regard $n$-strings with $n$
of order $N$, and not the fundamental string over which is
supposed to be built the dual string theory. So the hadrons that
becomes a bag are the exotic mesons $q^n$-$\overline{q}^n$ with $n
\sim N$.\footnote{This should also overcome previous problems
founded in trying to relate large $N$ QCD and bag models
\cite{Hansson}.}

The string tensions $T_V(n)$ has been much investigated, both
theoretically and numerically (see \cite{Greensite:2003bk} for a
review), in particular the ratio of string tensions defined by
$R(n,N)=\frac{T_V(n)}{T_V(1)}$. From the theoretical side there
are two important predictions: the Casimir scaling and the sine
formula:
\begin{description}
\item[Casimir scaling.] In an intermediate range of distance, the
force between two static sources is proportional to the quadratic
Casimir of the representation
\cite{Ambjorn:1984mb,DelDebbio:1996xm}. As $N$ becomes large, this
range goes to infinity and the extrapolation suggest the ratio for
string tensions \beq R(n,N) = \frac{n(N-n)}{N-1} \ .\eeq This
procedure is not rigorous, since we should first make the distance
to go to infinity, and then make the large $N$ limit. It has be
noted \cite{Armoni:2003ji} that the Casimir scaling, when $n$ is
kept fixed and $N$ goes to infinity, has corrections $\frac{1}{N}$
instead of the expected $\frac{1}{N^2}$. This should rule out the
Casimir scaling but there are also different opinions on that
\cite{Teper:2002kh}. \item[Sine formula.] This formula first
appeared first in \cite{DS} \beq R(n,N) =
\frac{\sin{\left(\frac{\pi
n}{N}\right)}}{\sin{\left(\frac{\pi}{N}\right)}} \ ,\eeq studying
$\N=2$ $SU(N)$ gauge theory, softly broken by the adjoint mass
term $\mu \Tr \Phi^2$. In \cite{Hanany:1997hr}, in the MQCD
contest, has been shown that this formula is true also in the
opposite limit $\mu \to \infty$. In \cite{Auzzi:2002dn} it has
been shown that the sine formula doesn't hold in the intermediate
regime but has non universal corrections.  The formula reappeared
in \cite{Herzog:2001fq} in the contest of cascading gauge
theories. Another meaning of the sine formula has appeared
recently in \cite{Gliozzi:2005en}. The conclusion is that the sine
formula, even if not directly derived in the QCD contest, present
some universal characteristics that makes it a good candidate for
QCD, or for any confining gauge theory.
\end{description}
 On the lattice side, a lot of works
have been done to compute the tensions in pure $SU(N)$ Yang-Mills
\cite{vicari,lucini}.  The most recent work on the subject
\cite{latest} gives results up to $N=8$. In the end of the paper
we will confront our theory with these results.

The paper is organized as follows. In Section \ref{firstsection}
we consider the wall vortex in the abelian Higgs model. First we
review the results of \cite{wallandfluxes} in the case where the
Coulomb phase has the same energy of the Higgs phase. Then we
extend the results to the most general case where the Coulomb
phase is a stationary point of the potential (this must always be
due to the $U(1)$ symmetry).
 In Section \ref{secondsection} we study a model in which there are $\mathbb{Z}_N$ solitonic magnetic strings that in the large $N$ limit become wall vortices.
  In Section \ref{thirdsection} we try to apply our ideas to
confining $\mathbb{Z}_N$ string in ordinary QCD. Even if our
reasoning doesn't follows directly from QCD first principles, it
gives a sharp prediction for the ratio of string tensions in the
large $N$ limit. The future lattice computations should easily
prove or disprove our hypothesis.

\section{The Abelian Theory and the Wall Vortex \label{firstsection}}

We consider the abelian gauge theory coupled to a charged scalar
field
\begin{equation}
\label{BasicVortex} {\cal L}=-\frac{1}{4 e^2}F_{\mu\nu}F^{\mu\nu}-
|(\de_{\mu}-i A_{\mu})q|^2-V(|q|)\ .
\end{equation}
In \cite{wallandfluxes} we considered the case in which the
potential has two degenerate vacua, one in the Coulomb phase and
the other in the Higgs phase. In \ref{abeliandeg} we briefly
review the results that we have obtained. In \ref{abeliannondeg}
we make a step forward trying to generalize the wall vortex idea
to the most general case, where the Coulomb phase is not a true
vacuum but only a stationary point.

\subsection{Coulomb-Higgs model and the surface bag \label{abeliandeg}}

\begin{figure}[h!t]
\begin{center}
\leavevmode \epsfxsize 9 cm \epsffile{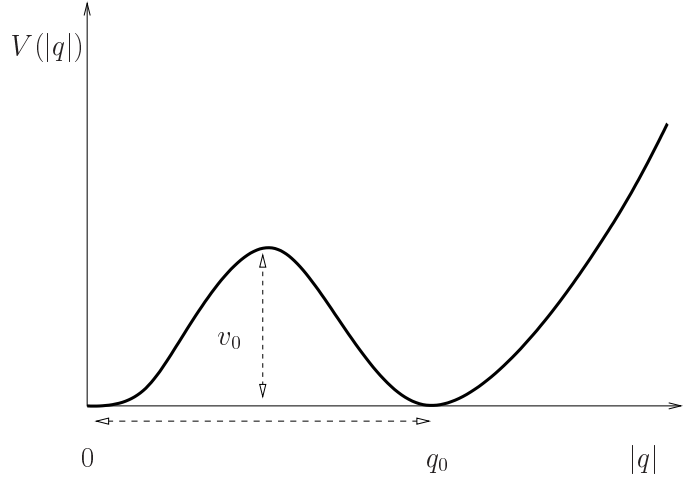}
\end{center}
\caption{\footnotesize A potential with two degenerate vacua:
$q=0$ is in the Coulomb phase while $|q|=q_0$ is in the Higgs
phase.} \label{potential}
\end{figure}

Now we consider the potential of Figure \ref{potential}. There are
two degenerate vacua: one in the Higgs phase where $|q|=q_0$, and
the other in the Coulomb phase. For this reason we call it abelian
Coulomb-Higgs model.
 The magnetic vortex
\cite{Abrikosov,NielsenOlesen} in the Higgs vacuum is nothing but
the  wall interpolating between the two vacua, compactified on a
cylinder and with the Coulomb vacuum left inside. The energy
density as function of the radius is \beq \label{qual}
T(R)=\frac{{\Phi_B}^2}{2 \pi e^2 R^2}+T_W 2 \pi R \ ,\eeq where
$\Phi_B$ is the magnetic flux. The magnetic flux is quantized in
integer values $\Phi_B=2\pi n$.  The stable configuration is the
one that minimizes the tension: \beq \label{eqscal} T_V=3
\sqrt[3]{2} \pi \; \left(\frac{n T_W}{e}\right)^{\frac{2}{3}} \ ,
\qquad R_V = \sqrt[3]{2} \; \sqrt[3]{\frac{n^{2}}{ e^{2} {T_W}}} \
. \eeq In this simple calculation we have neglected the thickness
of the wall $\Delta_W$ and this is the crucial point.  We can
trust (\ref{eqscal})  only if the radius if the vortex $R_V$ is
much greater than $\Delta_W$. In \cite{wallandfluxes} we argued
that (\ref{eqscal}) can be used in a self-consistent way to
determine its region of validity. Taking $R_V = \sqrt[3]{2} \;
\frac{n^{\frac{2}{3}}} {e^{\frac{2}{3}} {T_W}^{\frac{1}{3}}}$ for
true, we increasing $n$ keeping fixed all the other parameters of
the theory.  Note that $\Delta_W$ does not depend on $n$. There
will be some value $n^*$ above which $R_V \gg \Delta_W$ and so
(\ref{eqscal}) can be trusted.

\subsection{Higgs model and the volume bag \label{abeliannondeg}}

\begin{figure}[h!t]
\begin{center}
\leavevmode \epsfxsize 9 cm \epsffile{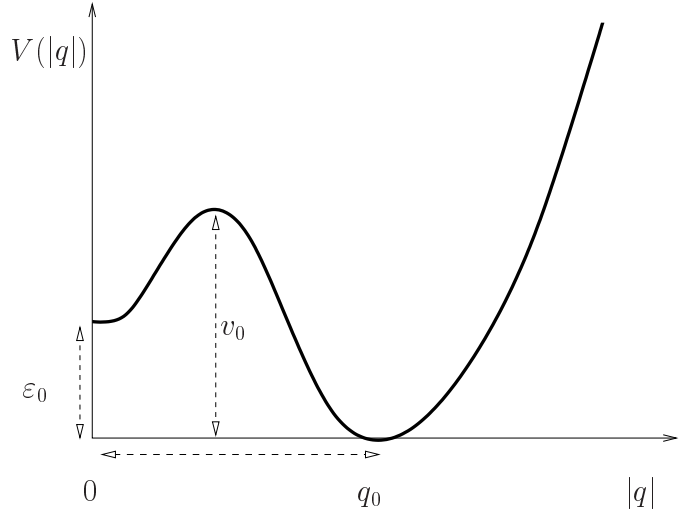}
\end{center}
\caption{\footnotesize  The vacuum $q=0$ is in the Coulomb phase,
it is metastable and its energy density is $\varepsilon_0$. The
other vacuum at $|q|=q_0$ is in the Higgs phase and has zero
energy density. When $\varepsilon_0 =0$ the potential becomes the
one of Figure \ref{potential} and when $\varepsilon_0=v_0$ it
becomes that of Figure \ref{potentialinst}.} \label{potentialmeta}
\end{figure}

Now we discuss an abelian gauge theory with a potential like
Figure \ref{potentialmeta}.  The Coulomb vacuum is metastable and
has energy density $\varepsilon_0$.

Also in this case is convenient to think of the flux tube as a
domain wall compactified in a cylinder, with the metastable
Coulomb vacuum inside and the true Higgs vacuum outside.
Neglecting the thickness of the wall, we can write the tension as
function of the radius:\footnote{To derive this formula we could
first think of the vortex tension as function of two parameters,
the radius $R$ and the thickness $\Delta$ of the shell where the
$q$ field goes from $0$ to $q_0$. The tension is roughly
$T(R,\Delta)=\frac{ n^2}{e^2 R^2} + \frac{{q_0}^2 R}{\Delta} +  R
\Delta v_0 + \varepsilon_0 R^2$. Minimizing with respect to
$\Delta$ we obtain $\Delta_W \sim \frac{q_0}{\sqrt{v_0}}$. So we
can interpret it  as  a domain wall since is independent on $n$
and $R$.} \beq \label{unavariabile} T(R) = \frac{2 \pi n^2}{e^2
R^2} + T_W 2 \pi R + \varepsilon_0 \pi R^2 \ .\eeq There are two
regimes in which (\ref{unavariabile}) can be easily solved.
\begin{description}
\item[Surface (or SLAC) bag.]  This region is when $n$ satisfies
the conditions \beq \label{surface} \frac{{q_0}^2 e}{ \sqrt{v_0}}
\ll n \ll \frac{{q_0}^2 e}{ \sqrt{\varepsilon_0}} \ . \eeq In this
limit the surface term in (\ref{unavariabile}) dominates over the
volume term and the minimization gives: \beq T_V=3 \sqrt[3]{2} \pi
\; \left(\frac{n {T_W}}{e}\right)^{\frac{2}{3}} \ , \qquad R_V =
\sqrt[3]{2} \; \sqrt[3]{\frac{n^{2}}{ e^{2} {T_W}}} \ . \eeq Note
that the surface region  (\ref{surface}) exists only if
$\varepsilon_0 \ll v_0$. As $\varepsilon_0$ is increased until it
reaches $v_0$, the SLAC region is eaten by the MIT region.
\item[Volume (or MIT) bag.] This region is when $n$ satisfies the
condition
 \beq \frac{{q_0}^2 e}{\sqrt{\varepsilon_0}} \ll n\ . \eeq  In this limit the volume term
in (\ref{unavariabile}) dominates over the surface term and the
minimization gives: \beq \label{volume} T_V = 2\sqrt{2 }\pi \;
\frac{n \sqrt{\varepsilon_0}}{e} \ , \qquad R_V = \sqrt[4]{2} \;
\sqrt{\frac{n}{ e \sqrt{\varepsilon_0}}} \ .\eeq Note that the
tension is proportional to $n$, as happens in the BPS case.
\end{description}

\subsubsection*{General conjecture}

\begin{figure}[h!t]
\begin{center}
\leavevmode \epsfxsize 9 cm \epsffile{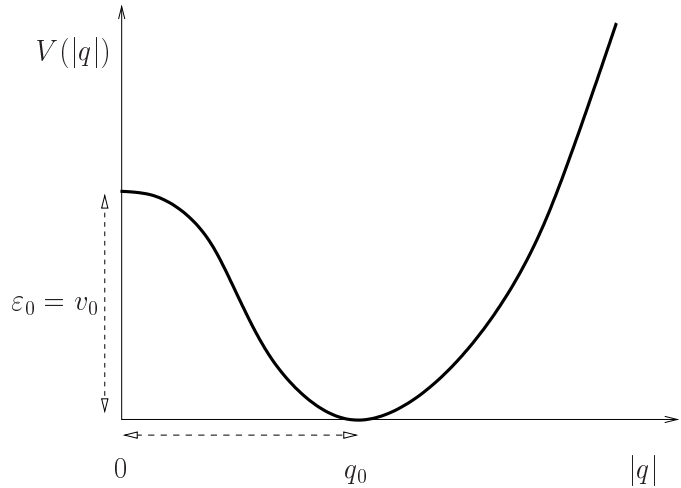}
\end{center}
\caption{\footnotesize  In the extreme case $\varepsilon_0=v_0$
the Coulomb vacuum is still stationary but instable.}
\label{potentialinst}
\end{figure}

The general conjecture is that the result (\ref{volume}) works for
every potential, even if the Coulomb phase is not metastable but
instable like in Figure \ref{potentialinst}.
Let's write the conjecture for clarity. {\it Consider the abelian
Higgs model (\ref{BasicVortex}) with a general potential that has
a true vacuum at $|q|=q_0\neq 0$ and a Coulomb phase with energy
density $V(0)=\varepsilon_0 \neq 0$. Call $T_V(n)$ the tension of
the vortex with $n$ units of magnetic flux. The claim is that \beq
\lim_{n \to \infty} T_V(n) = 2\sqrt{2 }\pi \; \frac{n
\sqrt{\varepsilon_0}}{e} \ .\eeq} We will give a substantial check
of this statement in \cite{Sven} using numerical computations.

\subsubsection*{A check}

Now we make a non trivial check of the conjecture using the famous
example solved by Bogomoln'y \cite{Bogomolny}. When the potential
is \beq \label{BPSpotential} V(|q|)=\frac{e^2}{2}(|\phi|^2-\xi)^2
\ ,\eeq the tension is \beq \label{BPStension} T_V=2 \pi n \xi
\eeq for every value of $n$. Solving the model with our trick, the
result must coincide with (\ref{BPStension}).  For the BPS
potential (\ref{BPSpotential}), the energy density of the Coulomb
vacuum is $\varepsilon_0=\frac{e^2\xi^2}{2}$ and, using
(\ref{volume}), we find exactly (\ref{BPStension}). This can
hardly be considered only a coincidence.

\section{ Solitonic $\mathbb{Z}_N$ Strings \label{secondsection}}

The principal aim of this paper is to see if it possible to have a
nonabelian generalization of the wall vortex. In this section we
consider a toy model of solitonic and magnetic $\mathbb{Z}_N$
strings. To have this we need an $SU(N)$ gauge theory with scalar
fields that are not charged under the center of the group. If in
some vacuum the scalar vevs brake completely $SU(N)$, then the
solitonic strings are stabilized by the homotopic group \beq
\label{homotopy}
 \pi_1\left(\frac{SU(N)}{\mathbb{Z}_N}\right)=\mathbb{Z}_N \ .\eeq

In the following we will consider the a broken version of the
$\N=4$ supersymmetric gauge theory. There are three adjoint scalar
fields that break completely the gauge group and are responsible
for the formation of the $\mathbb{Z}_N$ strings.

\subsection{$\N=1^*$ and the surface bag \label{exampletwo}}

The first case that we consider is the $\N=1^*$ $SU(N)$ gauge
theory, that is $\N=4$ broken to $\N=1$ by mass terms for the
chiral superfields $\Phi_1$, $\Phi_2$ and $\Phi_3$. The
superpotential is the $\N=4$ interaction plus the mass terms:
 \beq \label{supern1} W_{\N=1^*}=\frac{N  }{{g}^2} \Tr \left(
\Phi_1\left[\Phi_2,\Phi_3\right]+\frac{m_1}{2}{\Phi_1}^2+\frac{m_2}{2}{\Phi_2}^2+\frac{m_3}{2}{\Phi_3}^2
\right) \ .\eeq The stationary equations are, a part from
numerical factors, the commutation relations of the $SU(2)$
algebra \cite{Vafa:1994tf,Donagi:1995cf}. For example, deriving
(\ref{supern1}) with respect to $\Phi_3$, we obtain
$[\Phi_1,\Phi_2]=-m_3 \Phi_3$. Making the rescalings \beq \Phi_1 =
i \sqrt{m_2 m_3} \tPhi_1\ , \quad  \Phi_2 = i \sqrt{m_3 m_1}
\tPhi_2\ , \quad \Phi_3 = i \sqrt{m_1 m_2} \tPhi_3\ , \eeq we
obtain exactly the $SU(2)$ algebra as equations of motion: \beq
\left[\tPhi_i,\tPhi_j\right]=i \epsilon_{ijk} \tPhi_k \ .\eeq The
vacua of the theory are obtained choosing a partition of $N$ \beq
\label{generalvacuum} \sum_{d=1}^{N} d\, k_d =N \ ,\eeq so that
the $N \times N$ matrices $\Phi_i$ are covered with spin
$\frac{d-1}{2}$ representations of the $SU(2)$ algebra. The gauge
group is classically broken  \beq SU(N) \to
\frac{\otimes_{d=1}^{N} U(k_d)}{U(1)} \ ,\eeq and  the vacua of
the theory are divided into two classes:
\begin{description}
\item[Massive Vacua.] These vacua are called massive because there
is a mass gap. For every divisor of $N$ we must cover the matrices
$\tPhi_i$ only with representation of the same dimensionality. In
this case the gauge group is classically broken to
$SU\left(\frac{N}{d}\right)$ and there are no $U(1)$ factors. The
$SU\left(\frac{N}{d}\right)$ group confines and there is a mass
gap. This vacua are in one to one correspondence with the possible
phases of a general confining gauge theory \cite{'tHooft:1981ht}.
The works \cite{Vafa:1994tf,Donagi:1995cf} showed that the
$SL(2,\mathbb{Z})$ duality of the original $\N=4$ theory, exchange
the massive vacua among them.\item[Coulomb Vacua.] When there are
at least two representations with different dimensionality, the
unbroken gauge group has at least one $U(1)$ factor. Since there
is no strong dynamics for the $U(1)$'s, they survive in the
infrared giving some massless particles.
\end{description}

Before proceeding to the central point, we need to show some
qualitative properties of the domain walls in the $\N=1^*$ theory.
What is really important for us in the $N$ dependence of the
tensions and of the thicknesses of the walls. If we assume that
the walls are BPS saturated, their tension is governed by the
difference of the superpotential in the two vacua
\cite{DvaliShifmanWall}. For a general vacuum
(\ref{generalvacuum}) the superpotential (\ref{supern1}) is
proportional to the sum of the Casimirs of the spin
representations \beq W \sim {\tm}^3 N^2 \left(\sum_{d=1}^N k_d
(d-1)(d+1)\right) \ ,\eeq  where for simplicity we have used
$\tm=\sqrt[3]{m_1 m_2 m_3}$. And so the superpotential,  goes like
$\sim N^4$. Choosing two generic vacua, the interpolating wall
scales like \cite{Dorey:1999sj,Dorey:2000fc}: \beq
\label{standardwall} T_W \sim N^4 \, {\tm}^3\ , \qquad \Delta_W
\sim \frac{1}{\tm} \ . \eeq There are some exceptions to this
scalings. Take for example the domain wall between the Higgs
vacuum and the Coulomb vacuum where the $N$ is partitioned into a
$N-1$ and a $1$ representation (call it $(N-1,1)$ for simplicity).
The leading terms in the superpotential  cancel each other and the
wall goes like:\footnote{This mechanism is very similar to what
operates in $\N=1$ SYM where the tension of the domain wall
between adjacent vacua goes like $\frac{1}{N}$ instead of the
expected $\frac{1}{N^2}$ \cite{BranesandQCDWitten}. This point has
been clarified in \cite{ShifmanWall} and \cite{Dvali}.  } \beq T_W
\sim N^3 \, {\tm}^3 \ , \qquad \Delta_W \sim  \frac{1}{N \, \tm} \
. \eeq

Now we make our claim. The solitonic $\mathbb{Z}_N$ strings in the
Higgs vacuum are made by a domain wall compactified on a cylinder
with a Coulomb vacuum inside.  When $N$ becomes large the
energetically favorite Coulomb vacuum is the one where $N$ is
partitioned into a $N-1$ and a $1$ spin representation.  For
sufficiently large $N$ the radius of the $n$-strings becomes much
larger that the thickness of the wall, and so that the wall vortex
condition (namely $\Delta_W \ll R_V$) is satisfied.

First of all we check that the wall vortex condition is satisfied.
The Coulomb vacuum of interest, has only one $U(1)$ factor, the
one generated by the matrix
$\mathrm{diag}\left(1,\dots,1,-(N-1)\right)$. This is the only
generator in the Cartan subalgebra that, when exponentiated,
passes through all the elements of the center of $SU(N)$. In
particular the $n$-string has charge \beq \label{coulombgenerator}
\frac{n}{N}\left(\begin{array}{cccc}
1&&&\\&\ddots&&\\&&1&\\&&&-(N-1)\\
\end{array}\right) \ .\eeq
The tension as function of the radius $R$ is \beq \label{mag} T(R)
\sim \frac{n^2 N}{R^2}+N^{3} {\tm}^3 R  \eeq and, minimizing with
respect to $R$, we obtain:
 \beq  \label{coulombvacuum} T_V \sim \sqrt[3]{n^{2} N^{7}} \, {\tm}^2\ , \qquad R_V \sim \left(\frac{n}{ N}\right)^{\frac{2}{3}} \, \frac{1}{\tm} \ .
\eeq Since $\Delta_W \sim \frac{1}{N}$, for $N$ sufficiently large
$R_V$ is much greater than $\Delta_W$ and the wall vortex
approximation works. The ratio of string tensions is thus: \beq
R(n,N)=\min{(n^{\frac{2}{3}},(N-n)^{\frac{2}{3}})} \ .\eeq

Now we consider the other point: why the  Coulomb vacuum $(N-1,1)$
should be preferred with respect to the others? Or again: why the
$\mathbb{Z}_N$ strings should all choose the same Coulomb vacuum?
We don't have a rigorous proof for these questions but only an
argument in favor of it.  For example consider the $n$-string. Its
flux could also be generated by the $U(1)$ of the Coulomb vacuum
$(N-n,n)$: \beq \label{matrice} \frac{1}{N}
\left(\begin{array}{cc}
n \mathbf{1}_{N-n}&\\&-(N-n) \mathbf{1}_{n}\\
\end{array}\right) \ .\eeq
Note that this $U(1)$ cannot reach all the elements of the center
of $SU(N)$ so, if this generator comes out to be energetically
favorite, we would have a different $U(1)$ inside every
$n$-string.  What we are going to do is to evaluate the tension
and compare it with the one obtained with the Coulomb vacuum
$(N-1,1)$. The tension as function of the radius is \beq
\label{nottheright} T(R) \sim \frac{n(N-n)}{R^2} + {\tm}^3 N^4 R \
.\eeq The flux term has an $\frac{n(N-n)}{N}$ from the trace of
the square of (\ref{matrice}), and a power $N$ from the 't Hooft
scaling. The domain wall is an ordinary soliton that scales like
(\ref{standardwall}). The minimization of (\ref{nottheright}) with
respect to $R$ gives:
 \beq \label{nottrust} T_V \sim  \sqrt[3]{n(N-n) N^{8}} \, {\tm}^{2}  \ , \qquad
 R_V \sim \sqrt[3]{\frac{n(N-n)}{N^4}} \, \frac{1}{\tm} \ .\eeq
Note that in this case the radius scales like $N^{-\frac{2}{3}}$
if we send $n$ to infinity like $N$. So  we can not apply our
approximation since  the thickness of the wall
(\ref{standardwall}) remains finite. Note that (\ref{nottrust})
would give a tension that scales like $N^{\frac{10}{3}}$ would be
greater than the other one (\ref{coulombvacuum}). But, as we said,
the wall vortex approximation doesn't work and we have to find
another way to estimate the tension. The minimization of
(\ref{nottrust}) would give $R_V \sim N^{-\frac{2}{3}}$ that is
much lesser than the thickness of the wall $\Delta_W \sim 1$. This
means that the scalar fields are spread all over the radius of the
vortex and so their kinetic energy goes like $\frac{N^4}{R_V}$ and
dominates over the magnetic energy term that goes like
$\frac{N^2}{R_V}$. This imply that the vortex tension is
essentially given by the minimization of the scalar field action
$N^4\left(\de \phi \de \phi + V(\phi)\right)$ and so it scales
like $N^4$. This shows that the $(N-1,1)$ vacuum gives a lighter
tension than the $(N-n,n)$ vacuum in the large $N$ limit.

\section{Large $N$ QCD \label{thirdsection}}

Now we try to see if there is any chance that the wall vortex
scenario is realized in large $N$ QCD.

First we recall some well established results regarding the QCD
string tension at large $N$.  Consider the interaction between two
fundamental strings. It has been shown in \cite{Armoni:2003ji}
that interaction vanishes like $\frac{1}{N^2}$. Thus the tension
of the $n$-string, when $n$ is fixed and $N$ becomes large, is
equal to $n$ times the tension of the fundamental string \beq
\label{knowfact} T_V(n)=n T_V(1) + O\left(\frac{1}{N^2}\right) \ .
\eeq If instead we keep $n$ of order $N$, the interactions are of
order $1$. In fact, there are $\binom{n}{2}$ ways of making a
fundamental interaction between any couple of strings, and so in
total we have an interaction of order $\binom{n}{2}
\frac{1}{N^2}$.

Another information we will use is that the tension of the
fundamental string remains of order $1$ as $N$ is increased. In
fact its tension determines the mass of the meson and, for large
$N$, these masses remains finite
\cite{wittenbarioni,Manohar:1998xv}.

Now we explore the possibility that the $\mathbb{Z}_N$ strings of
QCD becomes wall vortices in the large $N$ limit with
$\frac{n}{N}$ kept fixed. A sort of magnetic dual of what happened
in the $\N=1^*$ theory, but with a Coulomb energy density
different from zero.
We write the tension as function of the radius of the vortex
considering only the dependence on $n$ and $N$: \beq \label{lento}
T(R) \sim \frac{n^2 }{N R^2}+N^{\alpha} \Lambda^4 R^2 \ . \eeq
Note that the energy of the electric flux term is $\frac{n^2}{N
R^2}$ instead of the $\frac{n^2 N}{R^2}$ obtained for the
solitonic magnetic vortex in (\ref{mag}). The reason is this:
normalizing the gauge kinetic term as we have done, $\frac{N}{g^2}
FF$, the magnetic monopoles have charge of order $1$ while the
electric particles have charge of order $\frac{g^2}{N}$ and this
brings the $N$ factor in the denominator of the electric field
energy density. The Coulomb energy density is instead $N^{\alpha}
\Lambda^4 R^2$ where $\Lambda^4$ is just for dimensional reasons
and $\alpha$ is a parameter that we are going to fix.

Minimizing (\ref{lento}) with respect to $R$ we obtain \beq
\label{match1} T_V \sim \frac{n}{\sqrt{N^{1-\alpha}}}\,\Lambda^2
 \ , \qquad R_V \sim \sqrt[4]{\frac{n^2}{N^{1+\alpha}}} \, \frac{1}{\Lambda} \ . \eeq
We could have a spectrum like (\ref{knowfact}), only if the
$\alpha$ parameter in (\ref{lento}) would be  equal to $1$. This
in fact is the only way to match (\ref{match1}) with
(\ref{knowfact}) since $T_V(1)$ is of order $1$. In this way the
tension is of order $n N^0$ and so, keeping fixed $n$ and sending
$N$ to infinity, we obtain a finite limit.

Now there are two left points to explain. Let us rewrite the
energy density, the tension and the radius now that we have fixed
$\alpha=1$: \beq \epsilon_0 \sim N^4 \, \Lambda \ , \qquad T_V
\sim n \,\Lambda^2
 \ , \qquad R_V \sim \sqrt{\frac{n}{N}} \, \frac{1}{\Lambda} \ . \eeq First
we have to explain how is possible to have a Coulomb energy
density that goes like $N \Lambda^4$. Second we have to explain
how the wall vortex approximation can work, that is we have to
find a domain wall between the confining vacuum and the Coulomb
vacuum with a thickness that goes like $\Delta_W \sim
\frac{1}{N}$. Both of these  points can be explained by an
effective Lagrangian that scales with an overall factor of $N^2$
\beq \label{generaleffectiveqcd} \L_{eff}=N^2 F[B,\de B ,\dots] \
,\eeq and a distance between the confining vacuum and the Coulomb
phase that is $\delta B \sim \frac{1}{N}$. Essentially is the same
mechanism that worked in Section
\ref{secondsection}.\footnote{It's also the same mechanism that
works for the pure $\N=1$ theory for domain walls between two
adjacent vacua \cite{BranesandQCDWitten,ShifmanWall,Dvali}. The
effective Lagrangian can be written as $\L_{eff}=N^2 [\frac{\delta
B}{\Delta}+\Delta]$. If $\delta B \sim \frac{1}{N}$, minimizing
with respect to $\Delta$ we obtain $\Delta_W \sim \frac{1}{N}$ and
$T_W \sim N$.}

\subsection{Lattice data and saturation}

Now we confront our theory with the lattice data. First of all we
make a brief discussion of the saturation limit. This is the best
way to confront the experimental results for different gauge
groups. Given any ratio of string tensions
$R(n,N)=\frac{T_V(n)}{T_V(1)}$, we rewrite it as a function of the
ratio $x=\frac{n}{N}$, and then rescale with a $\frac{1}{N}$
factor. For example the Sine formula and the Casimir formula
become respectively: \beq \frac{1}{N}\frac{\sin{\left(\frac{\pi
n}{N} \right)}}{\sin{\left(\frac{\pi}{N}\right)}} \longrightarrow
\frac{1}{\pi} \sin{(\pi x)} +O\left(\frac{1}{N^2}\right)\eeq \beq
\frac{1}{N}\frac{n(N-n)}{N-1} \longrightarrow x(1-x)
+O\left(\frac{1}{N}\right) \ .\eeq Note that in this way we can
plot $\frac{1}{N} R(n,N)$ with respect to $x$ in the same graph
for all the values of $N$. The saturation limit is when the
$\frac{1}{N}$ corrections can be neglected. This also the limit in
which the various data becomes dense and describe a continuous
curve. The Sine formula saturates to $\frac{1}{\pi} \sin{(\pi x)}$
while the Casimir formula saturate to $x(1-x)$. An important thing
to note is that both the Sine and the Casimir formula saturate
from above, that is the $o(\frac{1}{N})$ corrections are positive
. Our formula is instead \beq \min{(x,1-x)} +
O\left(\frac{1}{N}\right) \ .\eeq Since the deviation from an
exact wall vortex is given by the wall thickness of order
$\frac{1}{N}$, is natural to have $\frac{1}{N}$ correction. It's
also natural to expect a negative $\frac{1}{N}$ correction and so
a saturation from below.

\begin{figure}[!t]
\begin{center}
$\begin{array}{c}
 \epsfxsize 11 cm \epsffile{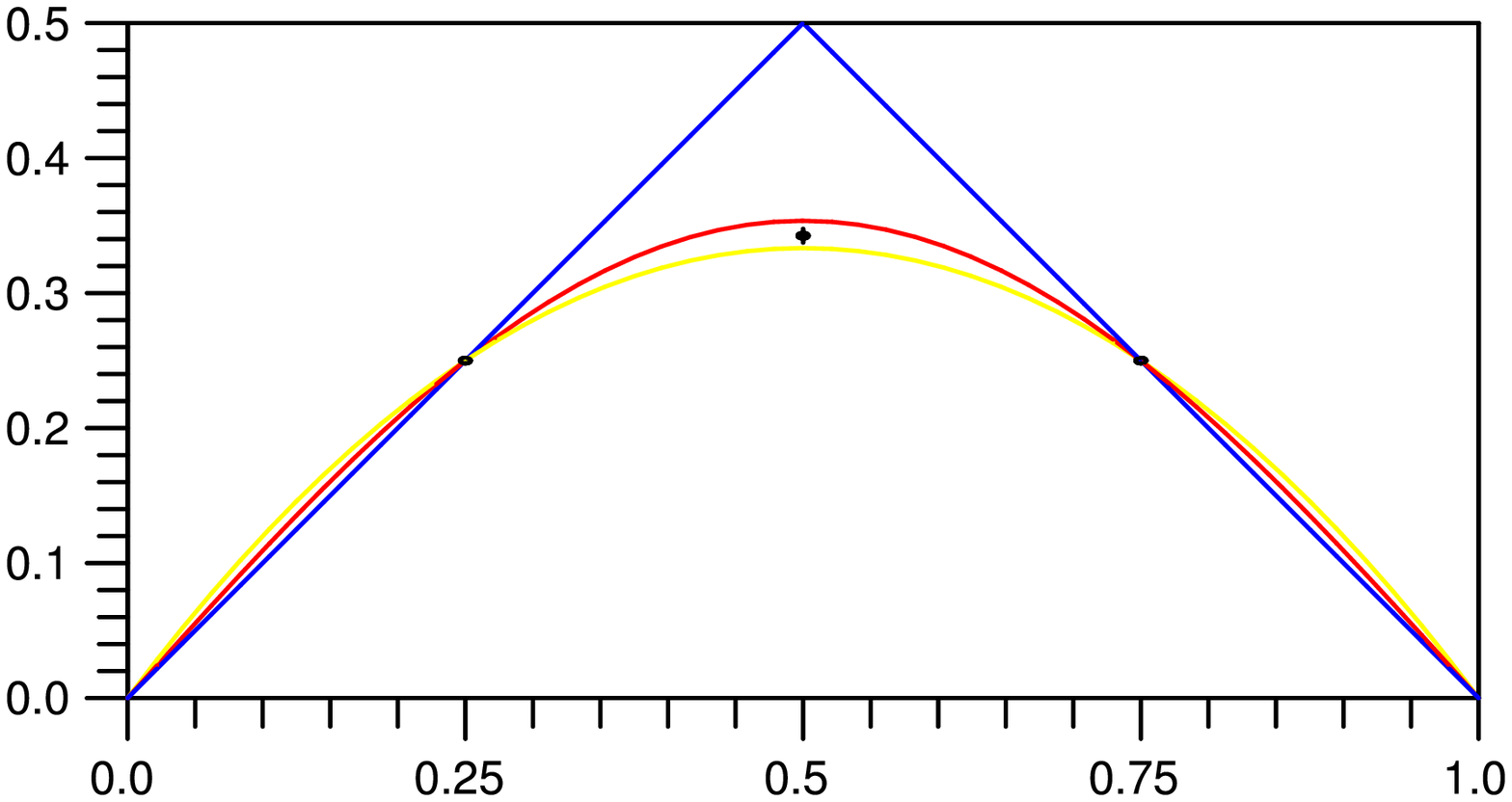} \\
\epsfxsize 11 cm \epsffile{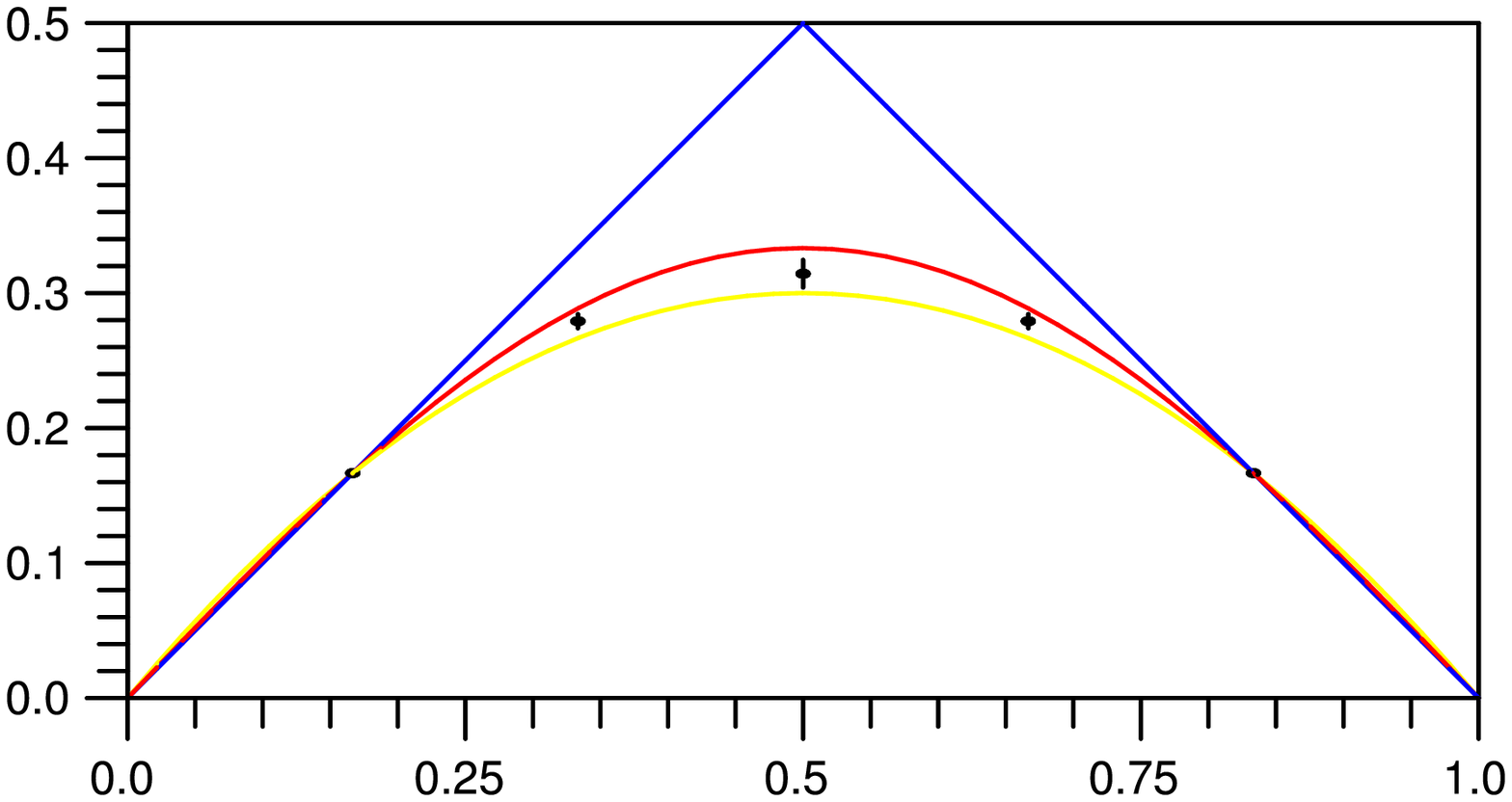}
\\ \epsfxsize 11 cm \epsffile{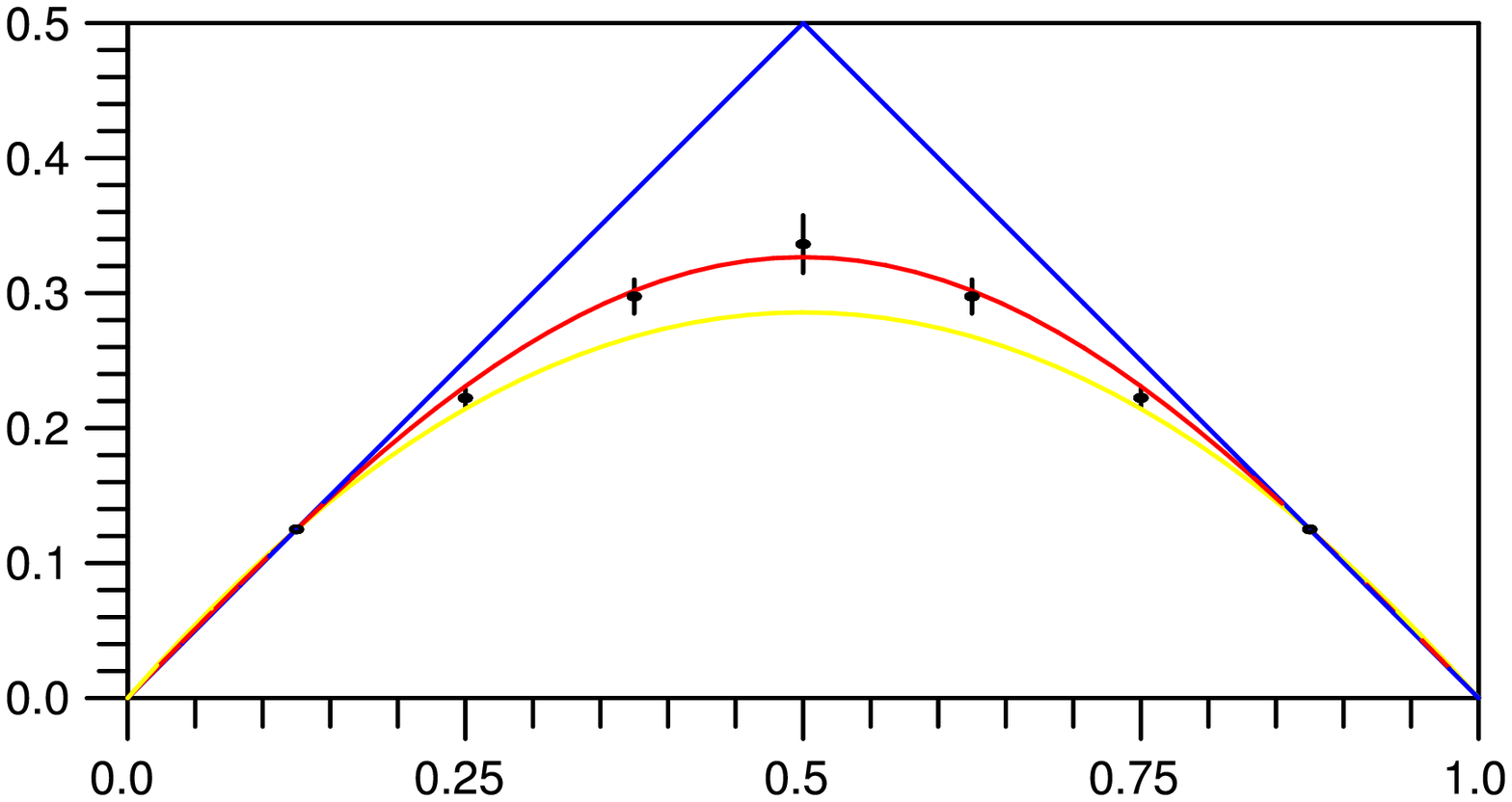} \\ \end{array} $
\end{center}
\caption{\footnotesize   In the three graph are reported the
lattice data taken from \cite{lucini}. They refer respectively to
$SU(4)$, $SU(6)$ and $SU(8)$. The lines plotted are respectively
the Casimir formula (green line), the Sine formula (red line) and
the $\min(x,1-x)$ formula (blue line).} \label{experiment}
\end{figure}

Finally we confront with the lattice experiments that are plotted
Figure \ref{experiment}. Up to now the largest $N$ for which
computations have been done is $N=8$.
 In principle these data could be
consistent with the formula $\min{(x,1-x)}$. To explain the
observed deviation we should have $\frac{1}{N}$ corrections with
coefficients of order $1$. When the saturation will be reached it
will be easy to prove or disprove our formula.

\newpage
\section* {Acknowledgement}

I thank Kenichi Konishi and Francesco Sannino  for useful
discussions. The work is supported by the Marie Curie Excellence
Grant under contract MEXT-CT-2004-013510 and by the Danish
Research Agency.

\end{document}